\newcommand{\gmas}{\mbox{$\dot{M_g}$}} 
\newcommand{\mlu}{\mbox{$M_{\odot}$\,yr$^{-1}$}}

\newcommand{\iras}{IRAS\,05280$-$6910}

\documentclass[useAMS,usenatbib]{mn2e}

\usepackage{txfonts}

\usepackage{graphicx}
\usepackage{longtable}

\title[LMC red-supergiants]
  {The mass-loss rates of red supergiants at low metallicity:
  Detection of rotational CO emission from two red supergiants in the Large Magellanic Cloud
  \thanks{{\it Herschel} is an ESA space observatory with science instruments provided by European-led Principal Investigator consortia and with important participation from NASA.}}
\author[Matsuura et al.]
{    Mikako Matsuura$^{1,2}$,
B. Sargent$^{3}$,
Bruce Swinyard$^{1,4}$\thanks{Prof. Bruce Swinyard passed away on 22nd of May 2015. He was very passionate about science and engineering, with good sense of humour. He made tremendous efforts to make space missions successful, including the {\em Herschel Space Observatory}. This paper is dedicated to his memory. We will miss him.}, 
Jeremy Yates$^{1}$, 
P. Royer$^5$,
               \newauthor 
M.J. Barlow$^1$,
Martha Boyer$^6$,
L. Decin$^{5,7}$,
Theo Khouri$^{7, 8}$,
Margaret Meixner$^{9,10}$, 
               \newauthor 
Jacco Th. van Loon$^{11}$, 
Paul M. Woods$^{12}$ \\
  $^1$ Department of Physics and Astronomy, University College London, Gower Street, London WC1E 6BT, UK\\
  $^2$ School of Physics and Astronomy, Cardiff University, Queen's Buildings, The Parade, Cardiff, CF24 3AA, UK \\
  $^3$ Center for Imaging Science and Laboratory for Multiwavelength Astrophysics, \\
           Rochester Institute of Technology, 54 Lomb Memorial Drive, Rochester, NY 14623, USA \\
  $^4$ RAL Space, Rutherford Appleton Laboratory, Chilton, Didcot, Oxfordshire, OX11 0QX, UK \\
  $^5$ Instituut voor Sterrenkunde, KU Leuven, Celestijnenlaan 200D B-2401, 3001 Leuven, Belgium \\
  $^6$ Observational Cosmology Lab, Code 665, NASA Goddard Space Flight Center, Greenbelt, MD 20771, USA \\
  $^7$ Astronomical Institute Anton Pannekoek, University of Amsterdam, PO Box 94249, 1090 GE Amsterdam, The Netherlands \\
  $^8$ Onsala Space Observatory, Dept. of Radio and Space Science, Chalmers University of Technology, 43992, Onsala, Sweden \\
  $^{9}$ Space Telescope Science Institute, 3700 San Martin Drive, Baltimore, MD 21218, USA \\ 
  $^{10}$ The Johns Hopkins University, Department of Physics and Astronomy, 366 Bloomberg Center, 3400 N. Charles Street, Baltimore, MD 21218, USA \\
  $^{11}$ School of Physical \& Geographical Sciences, Lennard-Jones Laboratories, Keele University, Staffordshire ST5 5BG, UK \\
  $^{12}$ Astrophysics Research Centre, School of Mathematics \& Physics, Queen's University, University Road, Belfast BT7 1NN, UK \\
  }
\date{Accepted 2016 July 22. Received 2016 July 22; in original form 2015 June 4}

\pagerange{\pageref{firstpage}--\pageref{lastpage}} \pubyear{2015}

\def\LaTeX{L\kern-.36em\raise.3ex\hbox{a}\kern-.15em
    T\kern-.1667em\lower.7ex\hbox{E}\kern-.125emX}

\begin{document}

\label{firstpage}

\maketitle

\begin{abstract}
Using the PACS and SPIRE spectrometers on-board the {\it Herschel Space Observatory}, we obtained spectra of
two red supergiants (RSGs) 
in the Large Magellanic Cloud (LMC).
Multiple rotational CO emission lines ($J$=6--5 to 15-14) and  15 H$_2$O lines were detected  from
IRAS\,05280$-$6910, and one CO line was detected from WOH\,G64.
This is the first time CO rotational lines have been detected  from evolved stars in the LMC.
Their CO line intensities are as strong as those of the Galactic RSG, VY\,CMa.
Modelling the CO lines and the spectral energy distribution results in an estimated mass-loss rate for 
IRAS\,05280$-$6910 of $3\times10^{-4}$\,$M_{\odot}$\,yr$^{-1}$.
The model assumes a gas-to-dust ratio and a CO-to-H$_2$ abundance ratio 
is estimated from the Galactic values scaled by the LMC metallicity ([Fe/H]$\sim-0.3$), i.e.,
 that the CO-to-dust ratio is constant for Galactic and LMC metallicities within the uncertainties of the model.
The key factor determining the CO line intensities and the mass-loss rate found to be the stellar luminosity.
\end{abstract}

\begin{keywords}
 (stars:) circumstellar matter --
 stars: mass-loss --
 stars: massive --
 ISM: molecules --
 (galaxies:) Magellanic Clouds --
 stars: AGB and post-AGB  --
  \end{keywords}

\section{Introduction}

At the end of their lives, stars lose a large fraction of their mass from their surfaces.
High-mass ($>8M_{\sun}$) stars evolve into the red-supergiant (RSG) phase,
and lose a large amount of mass through stellar winds. 
RSGs are considered to be the major  progenitors of type II-P supernovae \citep{Smartt:2009kr}.
High-mass stars lose a large amount of mass during the RSG phase,
determining the RSG mass at the time of a SN explosion \citep{Meynet:2015cq}.
In order to fully understand post-main sequence stellar evolution, it is important to characterise the mass loss  processes of RSGs.

The hydrodynamical simulation of RSG mass-loss is not yet fully modelled theoretically, but the basic mechanisms are believed to be the same as for asymptotic giant branch (AGB) stars, which are their lower-mass evolutionary counterparts. 
AGB stars are more populous than RSGs and thus better investigated.
It is widely accepted that stellar winds from RSGs and AGB stars are dust driven: 
stellar pulsations elevate some gas from the stellar surface, and from the levitated atmosphere dust grains are condensed 
\cite[e.g.][]{Hoefner:1997tv}.
The dust grains experience radiation pressure from the central star, triggering outward movement.
The dust motion drags the surrounding gas, driving stellar winds of both gas and dust.

Due to this dust-driven mechanism, the mass-loss rates from RSGs  and AGB stars 
should depend on two major parameters:
the {\bf luminosity} of the central star, and the {\bf metallicity} of the galaxy.

The luminosity of the central star determines the radiation pressure on the dust grains.
\citet{Wachter:2008p20280} predicted that the mass-loss rate should correlate with luminosity.
While Galactic RSGs have large uncertainties in measuring their luminosities,
extragalactic RSGs have well determined luminosities, as the distance to the stars 
can be adopted as the distance to the galaxy.
 With its close distance of 50\,kpc  \citep{Pietrzynski:2013ck}, the Large Magellanic Cloud (LMC) 
 can be used to study the impact on luminosity of mass-loss rates.

Secondly, the mass-loss rate is expected to approximately correlate with metallicity, which can be represented by the metallicity of the parent galaxy.
RSG mass loss is driven by radiation pressure on dust grains \citep[e.g.][]{Justtanont:1992p8184}. 
Dust grains from oxygen-rich stars are composed of metals, such as Fe, Si, O, Al, hence, a lower metallicity should result in a smaller mass of dust grains.
For a lower dust mass, the integrated cross-section for radiation pressure on dust grains should reduce, resulting in a lower mass-loss rate at least for oxygen-rich stars \citep{Bowen:1991p25238}.
Indeed, \citet{Marshall:2004p9396} measured the expansion velocities of LMC RSGs, using OH masers, and found lower expansion velocities for LMC RSGs compared to their Galactic counterparts, suggesting a metallicity  effect on the dust radiation-pressure driven mass-loss rate.

To date, studies of mass-loss rates in LMC RSGs have been based on their IR spectral energy distributions \cite[e.g.][]{vanLoon:1999wl, Groenewegen:2007p38, 2010ApJ...716..878S, Riebel:2012eq}, measuring their dust mass-loss rate, whereas historically, Galactic AGB stars and RSGs have been studied more intensively, using CO rotational lines \citep[e.g.][]{Knapp:1985p11494, Ramstedt:2008p25624}.
Complementary studies of CO emission in the LMC RSGs are important.

With the aim of investigating the physics and chemistry of the circumstellar envelopes of evolved stars at low metallicity, we have measured molecular emission lines at far-infrared (IR) and submillimeter wavelengths, including CO  rotational emission from two IR RSGs in the LMC.
Additionally, H$_2$O lines have been detected.
This is the first detection of CO rotational emission lines from RSGs in the LMC to our knowledge.
We report the analysis of these molecular lines.

\section{Observations and data reduction}

We observed two RSGs  with the {\em Herschel Space Observatory} \citep{Pilbratt:2010p29312}.
The primary target, \iras\, was chosen because it is the brightest LMC RSG at far-infrared wavelengths \citep{Boyer:2010ht},
with a 250\,$\mu$m flux of 205\,mJy.
Its mid-infrared spectrum shows 10\,$\mu$m silicate band in absorption with a peak of the SED at about 25\,$\mu$m.
OH and H$_2$O masers have been detected \citep{Wood:1992p8815, vanLoon:2001bt}. 
\iras\, was found towards the cluster NGC\,1984  \citep{vanLoon:2005fp}.

The second target is WOH\,G64, a well-known OH/IR star in the LMC \citep{Elias:1986p2265}.
This star has a strong silicate absorption in its mid-infrared spectrum, accompanied by OH masers \citep{Elias:1986p2265, Roche:1993p13250, Wood:1986p13258}.
Based on mid-infrared interferometric observations, \citet{Ohnaka:2008p19939} suggested the presence of a face-on torus around WOH\,G64.
This is one of the most luminous mass-losing RSGs in the LMC, with a bolometric luminosity of $\sim-9$\,mag \citep{vanLoon:1999wl, Levesque:2009ep}

The submillimeter spectrum of \iras\, was obtained, using the SPIRE Fourier Transform spectrometer
\citep{Griffin:2010p29303, Swinyard:2010p29297}
on board the {\em Herschel Space Observatory}.
The data were acquired on 2012 June 16th and 17th (OBSID: 1342247097 and 1342247098)
in a {\em Herschel} Open Time Program (OT1$_-$mmatsuur$_-$1).
The total duration of the observation was 13752\,s$\times$2.
The SPIRE FTS simultaneously covers the  short wavelength band 
(SSW; SPIRE Short Wavelength Spectrometer Array; 190--313\,$\mu$m;  957--1577\,GHz) and long 
wavelength band (SLW; SPIRE Long Wavelength Spectrometer Array; 303--650\,$\mu$m; 461--989\,GHz), 
yielding a frequency coverage from 444 to 1540\,GHz,
with a spectral resolution of 1.2\,GHz.
The FTS sensitivity varies across the spectra, with the best sensitivity available
 region at about 700--800 GHz  \citep{Swinyard:2014vy}, corresponding to the CO $J$=6--5 and 7--8 lines.
Unfortunately, the SLW spectrum suffered from a temperature drift in the SPIRE detectors, so that the continuum level is unreliable.
We discorded  the spectrum below $\sim$700 GHz, and  added some uncertainties to the SLW spectra above $\sim$700 GHz.
The spectrum was reduced in {\sc HIPE} version 11 \citep{2010ASPC..434..139O}.

The PACS \citep{Poglitsch:2010bm}
spectrum of \iras\, was obtained on 19th March 2013 (OBSID: 1342267866),
in a {\em Herschel} Open Time Program (OT2$_-$bsargent$_-$1).
The program targeted the 173\,$\mu$m CO $J$=15--14 and 186\,$\mu$m CO $J$=14--13 lines.
The target  CO lines were covered by the first-order grating with a resolving power of $R\sim1500$, 
while the second-order grating covered the 92\,$\mu$m H$_2$O $6_{43}$--6$_{34}$ line, with a resolving power of $R\sim2000$.
The chopped-nodded PACS range spectroscopy mode was used.
The total duration of the observation was 2240\,s.

 PACS obtained the spectrum of WOH\,G64 on 2013 April 12th (OBSID: 1342269921),
as part of the same programme (OT2$_-$bsargent$_-$1).
The spectral setting was the same as that for \iras, and
the total observing time was 6712\,s.

The PACS spectra were reduced with {\sc hipe} version 13. 
The absolute flux calibration was performed relative to the telescope background. 
Both targets were considered point-like, i.e. the fluxes were extracted from the central `spaxel' only, incling proper beam correction. 
Calibration set 65 was used for all reduction and calibration steps.

\begin{figure*}
\centering
  \resizebox{\hsize}{!}{\includegraphics{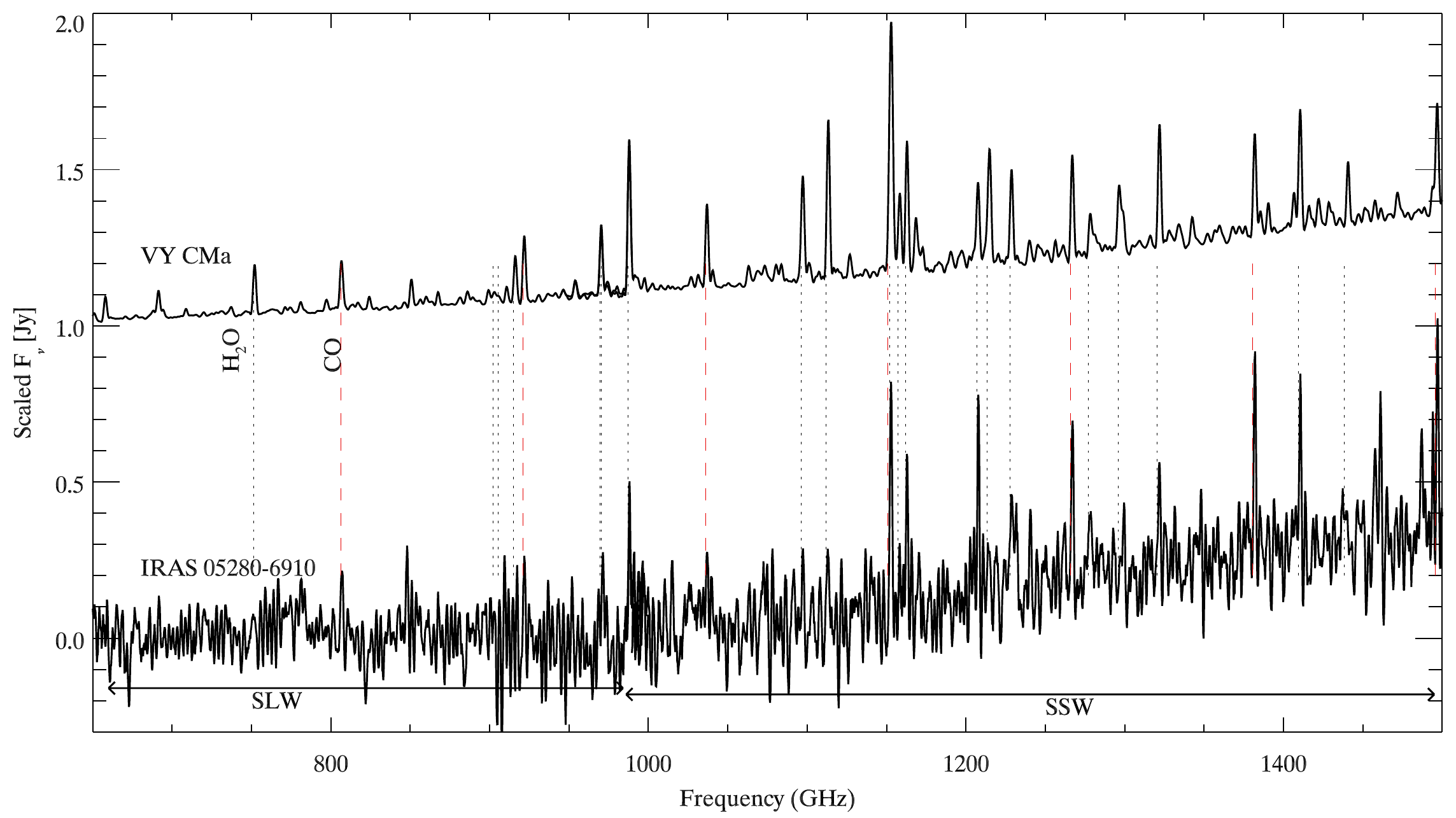}}
\caption{The SPIRE FTS spectra of the LMC red-supergiant, IRAS\,05280-6910, 
compared with the SPIRE FTS spectra of the Galactic red-supergiant, VY\,CMa.
The spectra were scaled for clarity.
The spectrum of \iras\, is unapodized, with the detected lines appearing as sharp peaks,
but with instrumental line shape of a sinc function included.
The spectrum VY\,CMa was apodized, i.e. instrumental line shape was removed, but the spectral resolution
was lowered.
\label{fig-fts-spectra}}
\end{figure*}

\section{Spectra and detected molecules}

Figure\,\ref{fig-fts-spectra} shows the SPIRE spectrum of \iras\,
in comparison to the spectrum of the Galactic red-supergiant VY\,CMa \citep{Royer:2010p29058, Matsuura:2013cz}.
CO rotational lines and some strong H$_2$O lines are detected, with their line intensities and identifications summarised in Table\,\ref{table-lines}.

\begin{figure*}
\centering
  \resizebox{\hsize}{!}{\includegraphics{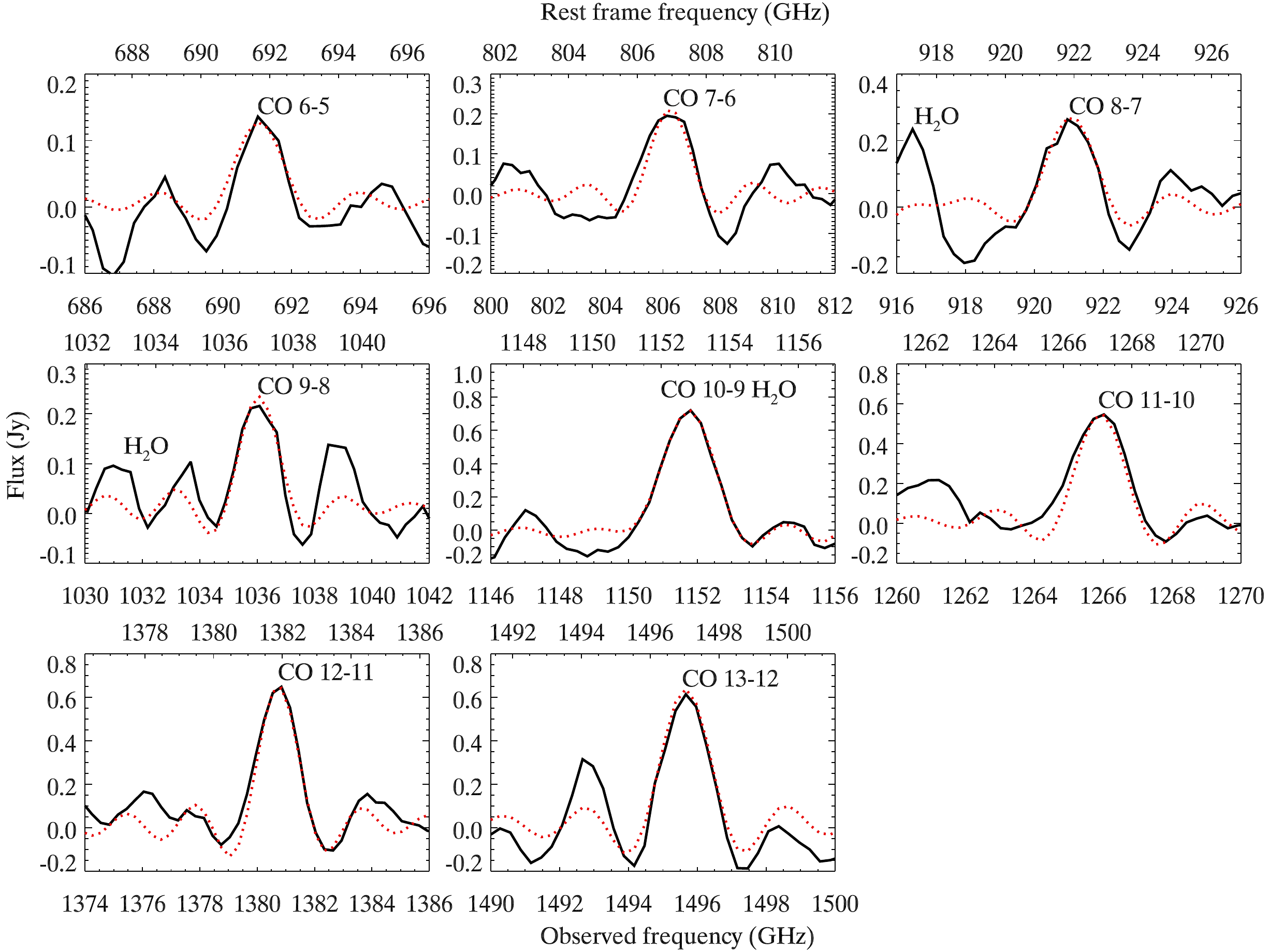}}
\caption{The SPIRE FTS unapodized spectra of the CO lines (continuous lines) of IRAS\,05280-6910, with instrumental model line profiles \citep[red dotted lines; ][]{2014SPIE.9143E..2DN}.
\label{fig-co-spectra}}
\end{figure*}

The detected CO lines are displayed in Fig.\,\ref{fig-co-spectra}.
The spectra were fitted using sinc instrumental line shape  \citep{2014SPIE.9143E..2DN}.
A systematic velocity of 270\,km\,s$^{-1}$  for the LMC
\citep{Marshall:2004p9396} was adopted.

\begin{figure*}
\centering
  \resizebox{\hsize}{!}{\includegraphics{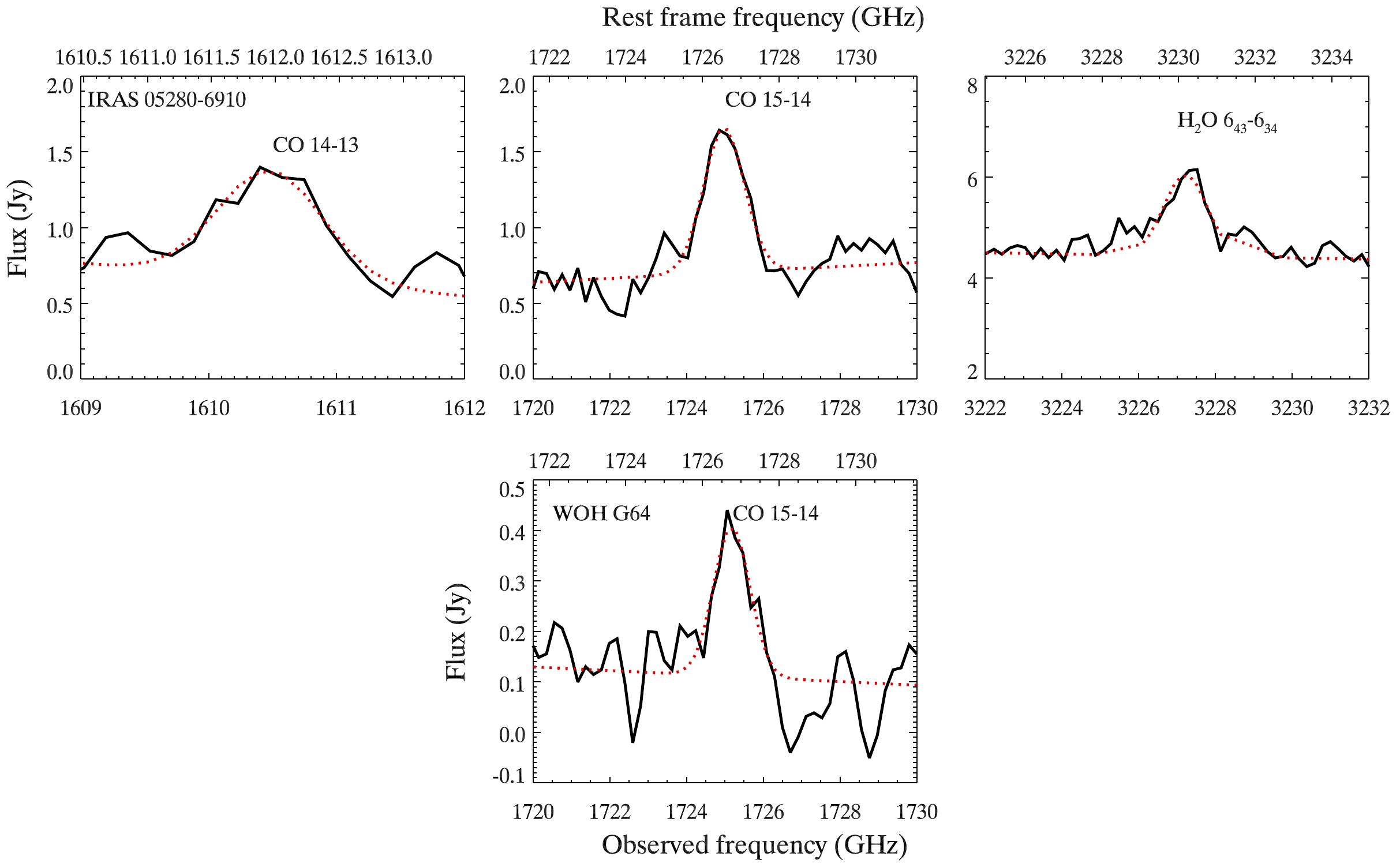}}
\caption{The PACS  spectra of IRAS\,05280-6910 (top) and WOH\,G64 (bottom). 
The red dotted lines show weighted Gaussians fitted to the measured lines.
\label{fig-pacs-spectra}}
\end{figure*}

Our PACS programme targeted two CO transitions ($J$=14--13 and 15--14) in the first grating order. 
As a bonus, the 92\,$\mu$m H$_2$O line was detected in the second order spectrum. 
Figure\,\ref{fig-pacs-spectra} displays the spectra of \iras\, and WOH\,G64, showing all three lines from \iras\, 
while only CO $J$=15--14 was detected from WOH\,G64.
 The lines observed by PACS were fitted by Gaussians with a quadratic underlying continuum, 
 allowing the line strength, line centre, line width, and coefficients of the quadratic continuum all to be free parameters.  
 These gaussians were fit to the lines plus surrounding long- and short-wavelength continuum, 
 except that the wings of the 92\,$\mu$m H$_2$O line of \iras\, were removed from the fitting process, 
 in order for the Gaussian to converge to a sensible fit.
The Gaussian fittings are plotted with red dotted lines in the figure, and the line intensities were estimated 
by integrating the Gaussian profiles.
The measured line intensities are summarised in Table\,\ref{table-lines-pacs}.

\begin{table*}
  \caption{ Lines in the SPIRE spectra of \iras.  \label{table-lines}}
\begin{center}
 \begin{tabular}{lrr  r@{$\pm$}lr@{$\pm$}lrrrr} \hline
Line name		& $\nu_0$	&	$\nu$	&  \multicolumn{2}{c}{  v}		&  \multicolumn{2}{c}{Int. Flux} \\ 
			& (GHz) & (GHz) 	& \multicolumn{2}{c}{(km\,s$^{-1}$)}	&  \multicolumn{2}{c}{($\times10^{-18}$\,W\,m$^{-2}$)}  \\ \hline 
\multicolumn{2}{c}{SLW (the SPIRE long wavelength band)} \\
CO $J$=6--5                      &  691.47 &  691.06  & 181  & 130  & 1.4  & 0.7 \\
CO $J$=7--6                      &  806.65 &  806.23  & 155  &  63  & 2.5  & 0.7 \\
p-H2O($\nu_2$=1,$3_{12}$--$2_{21}$)             &  902.61 &  902.22  & 129  & 118  & 1.3  & 0.7 \\
p-H2O(9$_{28}$--$8_{35}$)                   &  906.21 &  905.41  & 262  &  94  & 1.6  & 0.7 \\
CO $J$=8--7                      &  921.80 &  921.05  & 244  &  44  & 3.2  & 0.7 \\
p-H$_2$O (2$_{02}$--1$_{11}$)    &  987.93 &  987.19  & 224  &  22  & 6.0  & 0.7 \\
\hline
\multicolumn{2}{c}{SSW (the SPIRE short wavelength band)} \\
p-H$_2$O (2$_{02}$--1$_{11}$)    &  987.93 &  987.14  & 240  &  40  & 3.6  & 0.7 \\
CO $J$=9--8                      & 1036.91 & 1036.09  & 238  &  54  & 2.6  & 0.7 \\
o-H$_2$O (3$_{12}$--3$_{03}$)    & 1097.36 & 1096.31  & 288  &  50  & 2.6  & 0.7 \\
p-H$_2$O (1$_{11}$--0$_{00}$)    & 1113.34 & 1112.05  & 349  &  51  & 2.5  & 0.7 \\
CO $J$=10--9                     & 1151.99 & 1150.95  & 271  &  75  & 4.6  & 1.4 \\
o-H$_2$O (3$_{12}$--2$_{21}$)    & 1153.13 & 1152.09  & 271  &  46  & 7.4  & 1.5 \\
o-H$_2$O (6$_{34}$--5$_{41}$)    & 1158.32 & 1157.23  & 282  &  59  & 2.2  & 0.8 \\
o-H$_2$O(3$_{21}$--3$_{12}$)     & 1162.91 & 1161.85  & 274  &  22  & 5.9  & 0.8 \\
p-H$_2$O(4$_{22}$--4$_{13}$)     & 1207.64 & 1206.74  & 223  &  15  & 8.2  & 0.7 \\
o-H$_2$O($\nu_2$=1, 3$_{12}$--3$_{03}$) & 1214.66 & 1213.36  & 322  &  50  & 2.4  & 0.7 \\
p-H$_2$O (2$_{20}$--2$_{11}$)    & 1228.79 & 1227.67  & 274  &  30  & 4.0  & 0.7 \\
CO $J$=11--10                    & 1267.01 & 1265.89  & 266  &  18  & 6.4  & 0.7 \\
o-H$_2$O (7$_{43}$--6$_{52}$)    & 1278.27 & 1277.03  & 289  &  40  & 2.9  & 0.7 \\
o-H$_2$O (8$_{27}$--7$_{34}$)    & 1296.41 & 1295.98  & 101  & 115  & 1.0  & 0.7 \\
o-H$_2$O (6$_{25}$--5$_{32}$)    & 1322.07 & 1320.60  & 332  &  27  & 4.2  & 0.7 \\
CO $J$=12--11                    & 1382.00 & 1380.75  & 270  &  14  & 7.5  & 0.7 \\
o-H$_2$O (5$_{23}$--5$_{14}$)    & 1410.65 & 1409.34  & 277  &  18  & 5.9  & 0.7 \\
CO $J$=13--12                    & 1496.93 & 1495.66  & 253  &  13  & 7.3  & 0.7 \\ \hline
\end{tabular}\\
$\nu_0$ : vacuum frequency, $\nu$ : measured frequency, v : velocity shift
\end{center}
\end{table*}
\begin{table}
  \caption{ PACS measurements of CO and H$_2$O in \iras\, and WOH\,G64 spectra  \label{table-lines-pacs}}
\begin{center}
 \begin{tabular}{lll r@{$\pm$}l  r@{$\pm$}ll rrrrr}  \hline
Line name &  $\nu_0$  & $\nu$ &   \multicolumn{2}{c}{v}  &  \multicolumn{2}{c}{Int. Flux }    \\ 
 & (GHz) & (GHz) & \multicolumn{2}{c}{(km\,s$^{-1}$)} & \multicolumn{2}{c}{($10^{-18}$\,W m$^{-2}$)}  \\ \hline
{\bf \iras } \\
CO $J$=14--13						& 1611.79 & 1610.50 & 242 & 12 & 7.4  & 1.7    \\ 
CO $J$=15--14						& 1726.60 & 1724.99 & 280 &  6 & 12.7 & 1.2   \\
H$_2$O 6$_{43}$--6$_{34}$ 			& 3230.14 & 3227.24 & 269 &  4 & 23.0 & 2.8    \\
{\bf WOH G64} \\
CO $J$=15--14						& 1726.60 & 1725.19 & 245 &  11& 3.5  & 0.7   \\  \hline
\end{tabular}\\
$\nu_0$ : vacuum frequency, $\nu$ : measured frequency,
v : velocity shift
\end{center}
\end{table}

\section{Analysis}

\subsection{Spectral Line Energy Distributions}

\begin{figure}
\centering
   \resizebox{\hsize}{!}{\includegraphics{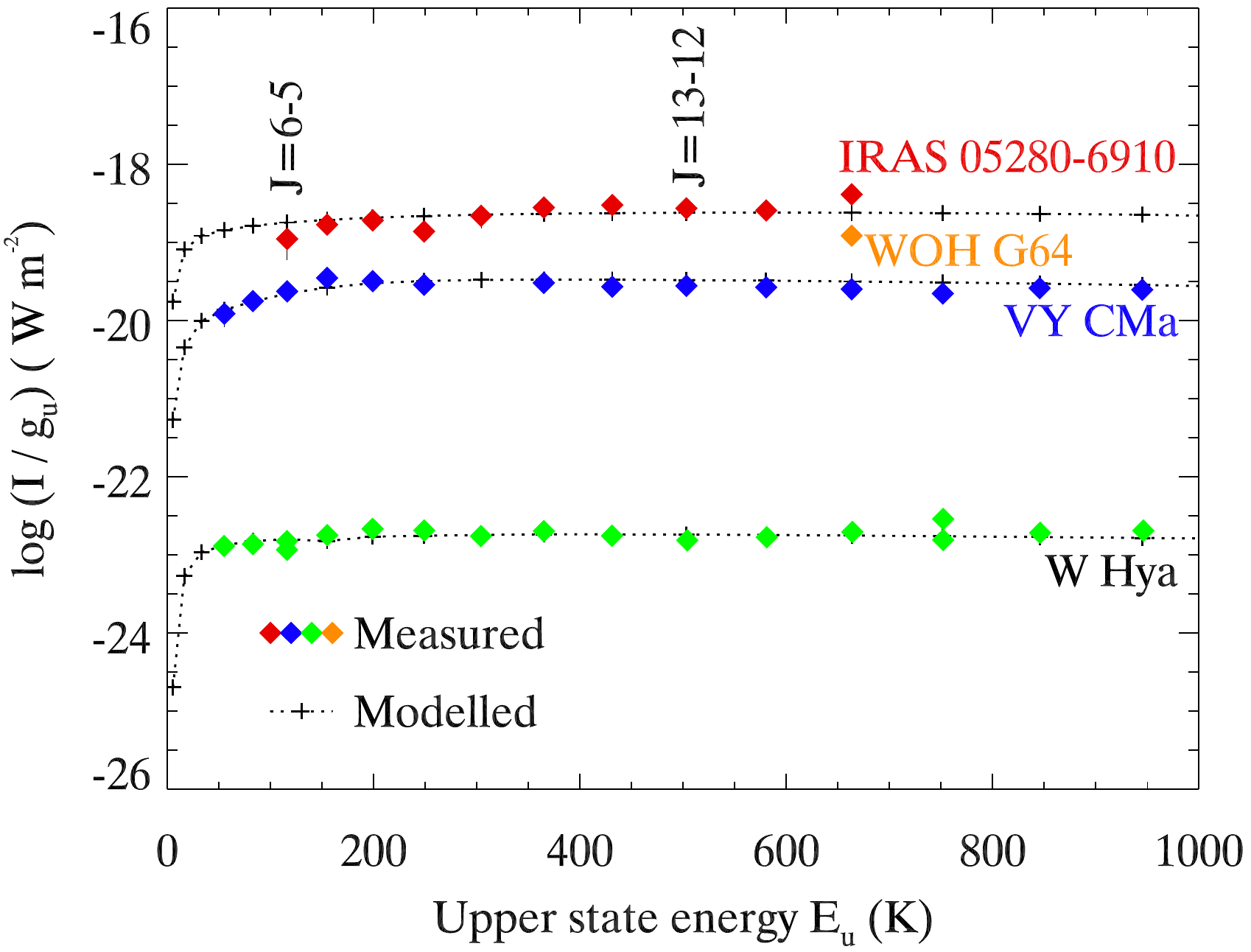}}
\caption{The spectral line energy distribution (SLED) for
\iras, VY\,CMa and W\,Hya (diamonds), with a single CO line measurement
for WOH\,G64. 
The SLEDs of VY\,CMa and W\,Hya were scaled to the
LMC distance,  by adopting their stellar distances of 1.14\,kpc and
78\,pc, respectively  \citep{Choi:2008p27073, Knapp:2003da}.
The non-LTE model fits are displayed with dotted lines.
\label{fig-fts-sled}}
\end{figure}

In order to examine the excitation temperature and column density
of CO, Spectral Line Energy Distributions (SLEDs) were used.
%
Fig.\ref{fig-fts-sled} shows the SLED of \iras.
The single CO line measurement of WOH\,G64 has also been plotted.
The figure also shows
{\em Herschel} measurements of SLEDs of the Galactic RSG VY\,CMa and the Galactic AGB star W\,Hya
 \citep{Matsuura:2013cz, Khouri:2014gz}.
The SLEDs of VY\,CMa and W\,Hya were scaled to the
LMC distance,  by adopting their stellar distances of 1.14\,kpc and
78\,pc, respectively  \citep{Choi:2008p27073, Knapp:2003da}.

\subsection{Non-LTE line radiative transfer modelling}

 In order to derive physical parameters from CO line intensities,
a non-local thermodynamic equilibrium (non-LTE) line radiative transfer code was used
to fit the SLEDs. As only \iras\, has multiple CO transitions detected,
we focus our modelling effort on this RSG.

 The first step was to construct a dust radiative transfer model to fit the spectral energy distribution (SED).
 This step provided some essential parameters for the CO modelling, such as
 dust temperature structure, which has an effect on the heating balance between  dust and gas collisions.
 Modelling the SED with {\sc dusty} provides the inner radius,
 the dust radial density and the temperature distribution of the dust envelope.
 These parameters were also converted into a dust mass-loss rate, and, hence, gas mass-loss rate,
 with an assumption of the gas-to-dust ratio.
 \citet{Boyer:2010ht} already presented modelling of the  SED of \iras\, 
 using {\sc 2-dust}  code \citep{Ueta:2003p10341}.
 We started by reproducing the {\sc 2-dust} results with {\sc Dusty}  \citep{Ivezic:1997tx}, 
as the {\sc Dusty} model is one of our default input models to CO modelling.
We used the  infrared flux measurements of  \iras\, collected by  \citet{Boyer:2010ht},
 including 3--24\,$\mu$m photometric data from {\em Spitzer} Magellanic Survey, SAGE \citep{Meixner:2006p2715},
the measurements of {\em Herschel} Magellanic Survey, HERITAGE \citep{Meixner:2013kr} covering 100--500\,$\mu$m,
and at 5--35\,$\mu$m a {\em Spitzer} IRS spectrum   \citep[SAGE-spec; ][]{Kemper:2010bw}.
 We used  $Ks$-band magnitudes from the VISTA Magellanic Cloud survey \citep[VMC; ][]{Cioni:2011iz}, prior to their public release of the magnitude of this star.
The PSF fitting resulted in a magnitude of 12.604$\pm$0.080 in the $Ks$-band.
It was converted to a Vega-system magnitude following the procedure described in \citet{Rubele:2015gy}, then converted to a flux in Jy.
The source was not detected in the VMC $z$ and $J$-bands.

An effective temperature of 3000\,K was adopted for \iras\, \citep{Boyer:2010ht}.
This is similar to the estimated temperature of WOH\,G64  and VY\,CMa
which have similar mass-loss rates of $\sim10^{-4}$\,\mlu. The effective temperature of VY\,CMa
has been estimated to be in the range of 2700\,K \citep{Monnier:2004dn} to 3650\,K \citep{Massey:2006p27317}.
\citet{Elias:1986p2265} estimated a spectral type of M7.5 for WOH\,G64, corresponding to an effective temperature
lower than 3450\,K \citep[=M5 ; ][]{Levesque:2005dx}, while \cite{Ohnaka:2008p19939} suggested to be 3200--3400\,K.
 \citet{Davies:2013ia} argued that the effective temperature of red-supergiants should be higher
than previously thought by $\sim$500\,K. Our experimental modelling with the effective temperature increased to 3500\,K 
resulted in a negligible ($<5\,\%$) increase in CO line intensities. 
We therefore adopted the 3000\,K effective temperature from \citet{Boyer:2010ht} but little difference in the CO modelling results for 
an effective temperature of 3500\,K.

\begin{figure}
\centering
  \resizebox{\hsize}{!}{\includegraphics{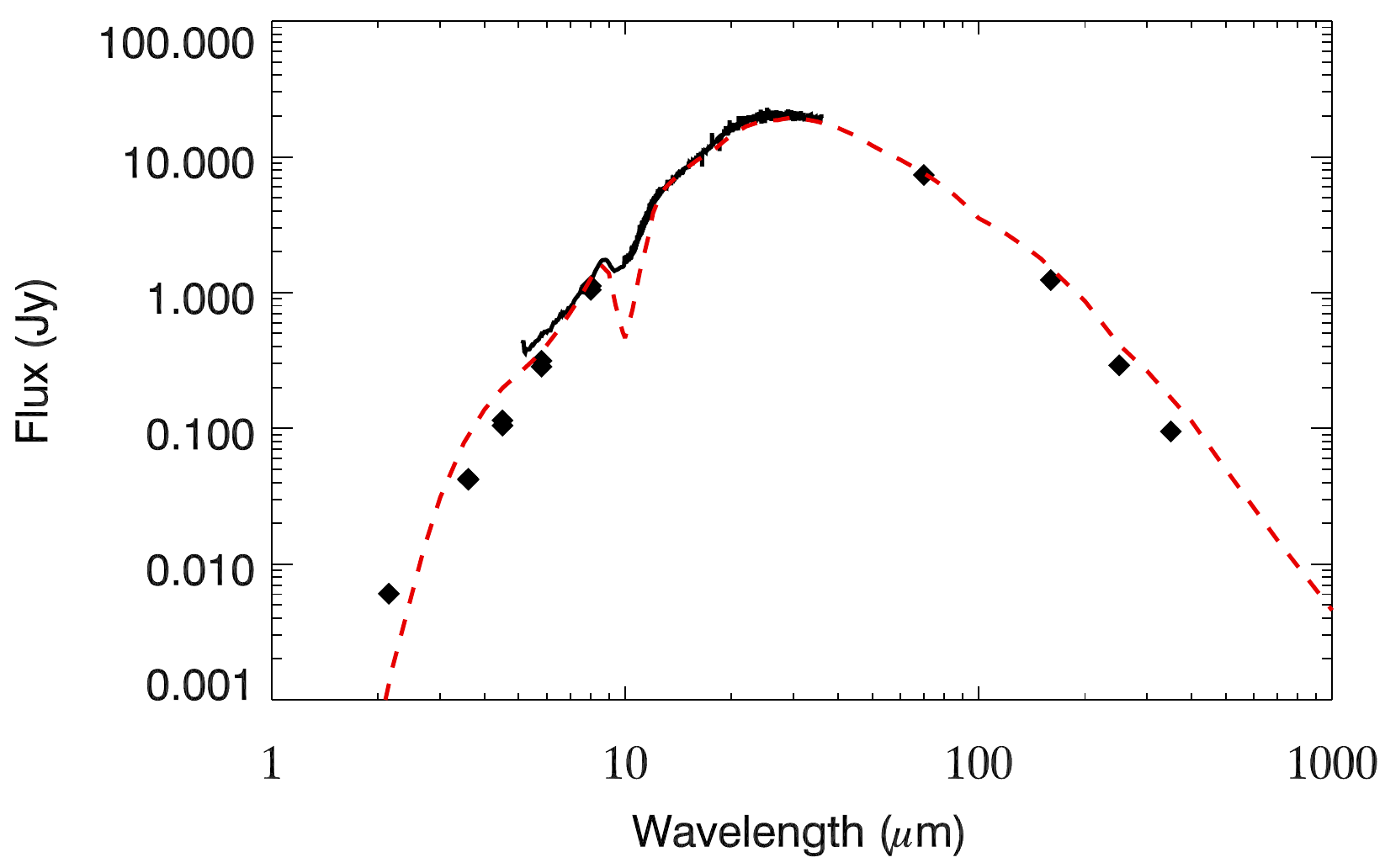}} 
\caption{The spectral energy distribution of \iras. 
The measured infrared fluxes are from the sauces described in the text.
The red dashed line shows the dust radiative transfer model fit to the SED.
\label{fig-fts-sed}}
\end{figure}

Our {\sc Dusty} fitted results for \iras\, are displayed in Fig.\ref{fig-fts-sed}.
The overall shape of the SED was fitted with silicate dust emission.
As found by \citet{Boyer:2010ht}, our modelling produced silicate absorption 
 stronger than that observed.  This could be due to the outflow being slightly asymmetric,
 as recently found for the Galactic RSG VY\,CMa  \citep{OGorman:2014uh}.
 However, currently we do not have any information about asymmetries for \iras,
 so we adopted symmetric modelling.
 Alternatively, \citet{Boyer:2010ht} suggested a potential contribution of silicate emission from a nearby
 RSG (WOG\,G347) onto the {\em Spitzer} IRS slit, but not visible at longer wavelengths. 
Our {\sc Dusty} parameters are  summarised in Table\,\ref{table-smmol},
which are consistent with those obtained from \citet{Boyer:2010ht}'s {\sc 2-dust} modelling.
 We evaluated the {\sc Dusty} input parameters with a $\chi^2$-analysis.
We found that $T_{\rm dust}$ and $\tau$ are coupled parameters,
and the uncertainty of the dust temperature is $\pm$50\,K, and that of the optical depth at 0.55\,$\mu$m ($log \tau_{0.55}$) is $\pm$0.1,
with increasing dust temperature requiring higher dust optical depth.
We used dust optical constants for amorphous silicates \citep{Ossenkopf:1992p27453},
and the MRN grain size distribution with an index of 3.5, and grain size of 0.005--1\,$\mu$m
\citep{Mathis:1977hp}.
 Assuming a gas-to-dust ratio of 500, which is expected from the abundance of refractory elements in the LMC ISM
 \citep{Gordon:2014tm},
 the SED modelling resulted in a gas-mass loss rate of $3\times10^{-4}$\,\mlu.
 That carries 30\,\% uncertainty, according to the {\sc Dusty} manual\footnote{www.pa.uky.edu/$\sim$moshe/dusty/}.

 The second part of our modelling was 
to reproduce the CO rotational lines using the non-LTE line radiative transfer calculation
and level population code, {\sc SMMOL}  \citep{Rawlings:2001p28230}.
The {\sc SMMOL} code adopts the accelerated lambda integration (ALI) scheme
\citep{Scharmer:1985p28621, Rybicki:1991p28622}
with an ability to solve the radiative transfer problems in optically thin and thick circumstellar envelopes.
Solving level populations requires cross sections for molecular collision, and we adopted 
the CO--H$_2$ cross sections calculated by \citet{Yang:2010bb},
with molecular data from \citet{Muller:2005p29905}.
The code has been used for modelling the SLEDs of CO from $J=$4--3 to 22--21, as well as for H$_2$O masers
in the Galactic RSG, VY\,CMa \citep{Matsuura:2013cz}.

The majority of the {\sc SMMOL} input parameters for \iras\,  were constrained from SED modelling work
prior to the CO modelling.
Integrating over the SED provided luminosity of $2.2\times10^5$\, $L_{\odot}$ \citep{Boyer:2010ht}.
As discussed above, an effective temperature of 3000\,K was  adopted for \iras\, \citep{Boyer:2010ht}. 
These two parameters yielded a radius for the star ($R_{\rm star}$) of  $1.21\times 10^{14}$\,cm (=1738\,$R_{\sun}$).
The {\sc Dusty} model provided the inner radius, radial density and temperature structure of the dust envelope,
as well as the optical depth ($\tau_{0.55}$).
The parameters used for the \iras\, non-LTE modelling are listed in Table\,\ref{table-smmol}.
Another important parameter is the gas terminal velocity, which
was adopted to be 17\,km\,s$^{-1}$, based on OH maser measurements \citep{Marshall:2004p9396}.
The majority of our  input parameters were already well constrained
from previous studies.

The CO--H$_2$ abundance ratio is affected by the metallicity.
 A chemical model for the oxygen-rich AGB stars show that
essentially all available C atoms are locked up to CO \citep{Willacy:1997p27138},
so that the C abundance with respect to H$_2$ gives the CO abundance.
Scaling with the solar-abundance values for carbon with respect to hydrogen
\citep[8.43; ][]{Asplund:2009p27448}, and taking
[Z/H] = $-0.3$ for the LMC, the estimated CO--H$_2$ abundance is $2.7\times10^{-4}$. 
Overall, the model predicts that the high-$J$ CO line intensities scaled with the CO abundance. i.e.
a metallicity effect is found for the high-$J$ CO line intensities.
 The adopted CO/H$_2$ abundance ratio was based on the C and O abundances of the Sun, scaled to the LMC metallicity,
however, C and O abundances of high-mass stars can be potentially  modified due to nuclear synthesis \citep{Meynet:2015cq}. The current work ignores this effect.

\begin{figure}
\centering
  \resizebox{\hsize}{!}{\includegraphics{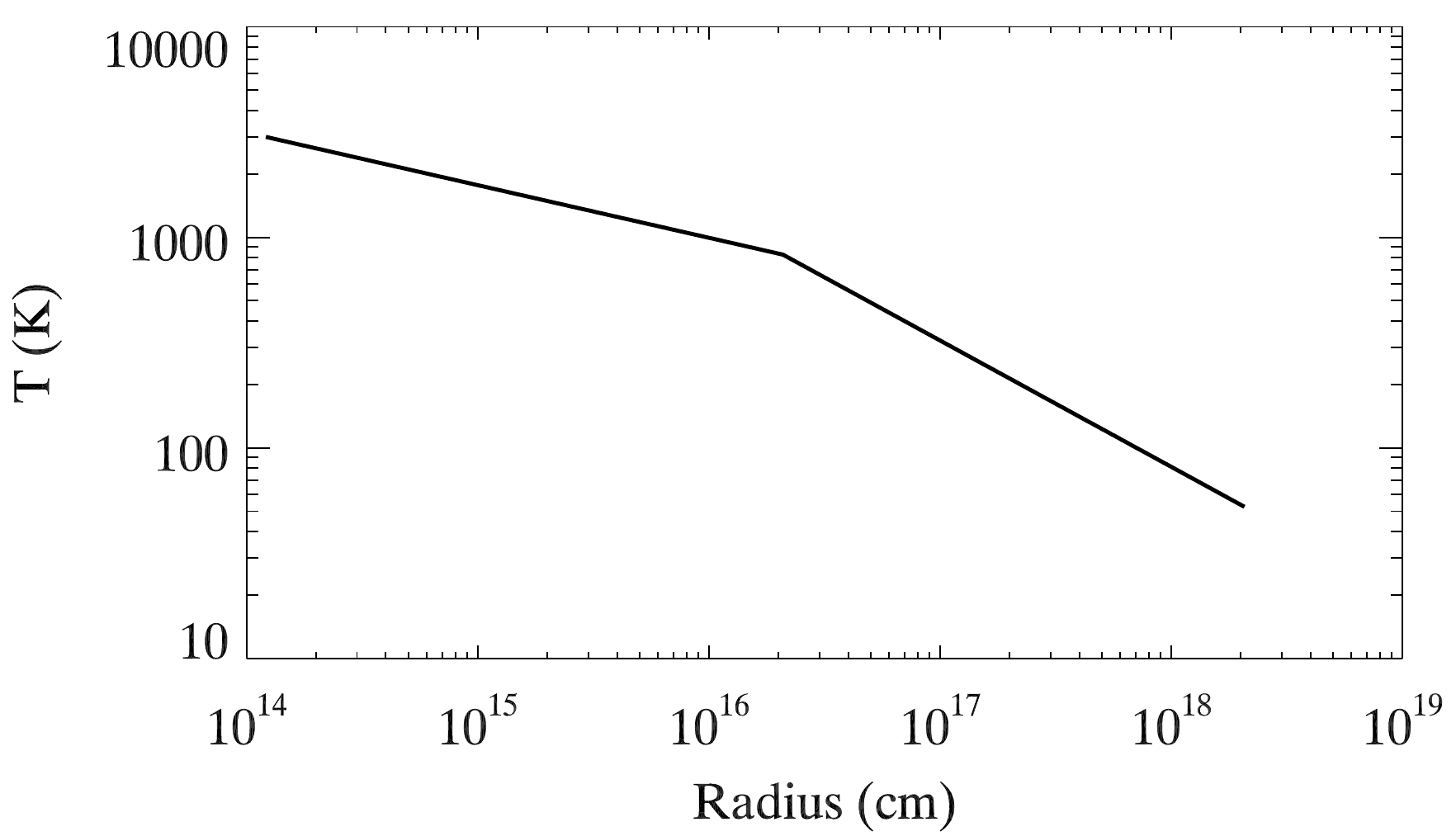}}
\caption{The temperature structure used for {\it SMMOL} modelling.
\label{fig-temp}}
\end{figure}

Three parameters were adopted from the  VY\,CMa modelling \citep{Matsuura:2013cz}:
the temperature and velocity structure gradient indices, and the turbulent velocity.
A  turbulent velocity of 1\,km\,s$^{-1}$  was adopted \citep{Decin:2006p27245, Matsuura:2013cz}.
The kinetic temperature gradient index 
$\alpha$, where $T_{\rm{kin}} \propto (r/R_{inner})^{-\alpha}$, has not yet been determined for circumstellar envelopes at low metallicity.
The balance between cooling and heating processes determines the temperature gradient.
Two major cooling processes are adiabatic cooling and  H$_2$O line emission \citep{Decin:2006p27245}.
The H$_2$O abundance has a metallicity dependence, as it is limited by the available oxygen abundance. 
The H$_2$O cooling is expected to be less efficient at low metallicity.  
On the other hand, key gas heating processes in the circumstellar envelope are associated with collisions with dust grains \citep{Decin:2006p27245}. 
At low metallicity, the dust-to-gas ratio is small; hence the heating rate is expected to be low, too.
Depending on the balance between these heating and cooling processes,
the temperature gradient can change slightly at low metallicity.
However, we found that changing $\alpha$ at $R>R_{\rm dust}$ from 0.5 to 0.6 made less than a few per cent difference
to the predicted CO line intensities. Such a small difference in the gas gradient in the circumstellar envelope is not crucial for CO modelling. 
We adopted of 0.6, the $\alpha$ value found for for VY\,CMa \citep{Matsuura:2013cz}, and   the adopted temperature structure is plotted in Figure\,\ref{fig-temp}.
Similarly, we examined $\alpha$ at $R<R_{\rm dust}$, by changing the value of VY\,CMa
(0.15) to 2.5. There was negligible difference in the predicted CO line intensities. The value of $\alpha$=0.25 was used
at $R<R_{\rm dust}$.

 The velocity structure parameter, $\gamma$
describes the velocity structure of the wind acceleration region. However,
the CO emitting region (approximately $r>100R_*$) is outside of the wind accelerating region, 
so  $\gamma$ is not important for the CO modelling.
Experimentally, when we changed $\gamma$ up to 5, the obtained CO line intensities remained unchanged,
because the velocity structure at $r>100R_*$ does not change.
Constraining $\gamma$ requires measurements of the velocity structure in the inner region.
That has been estimated for Galactic red-supergiants \citep{Decin:2010p29883}, suggesting $\gamma$ to be less than 1.
We adopted $\gamma=0.2$, following our modelling of VY\,CMa \citep{Matsuura:2013cz}.


 A density ($\rho$) law index $\beta=2$, where $\rho\propto(r/R_{inner})^{-\beta}$,  
 is adopted for the gas with a constant expansion velocity. Although this is not the case near the velocity accelerating
region, because this inner part does not affect the observed CO line intensities, it is not a major issue.

The outer radius ($R_{\rm out}$) was chosen to be large enough that predicted 
CO line intensities were unaffected by $R_{\rm out}$. 
As $J$=6--5 and  higher CO transitions
have their line forming regions much closer to the central star than low-$J$ CO lines (1--0 or 2--1)  \citep{Khouri:2014gz}, 
these high-$J$ line intensities are less sensitive to the choice of $R_{\rm out}$.
The $R_{\rm out}$ is the value corresponding to the dissociation radius of CO due to ISM UV radiation.
 In the LMC, a higher ISM UV radiation field is expected than for the Galactic ISM, due to  low dust extinction.
Hence, UV photons might  penetrate further into the circumstellar envelope,
dissociating CO at smaller radii in the LMC RSGs than for the Galactic counterparts \citet{McDonald:2014jr}.
Fortunately, such a low metallicity effect would not be important for the high-$J$ CO line intensities.

 In summary, although there are several parameters in {\em SMMOL} modelling,
key input parameters are the gas mass-loss rate and CO/H$_2$ abundance ratio, which determine overall height of CO line intensities,
and temperature index $\alpha$ that manipulates the tilt of the  CO SLED.

\begin{table}
  \caption{ Model parameters for \iras\,\label{table-smmol}}
\begin{center}
 \begin{tabular}{ll rrrrrr}
 \hline
 {\bf {\sc Dusty }} \\
 $T_{*}$ (K)					& 3000 \\
 Dust							& Silicate \citep{Ossenkopf:1994tq}  \\
 $T_{\rm dust}$ (K)				& 350$\pm$50  \\
\gmas (\mlu)					&  $3\pm1\times10^{-4}$  \\
gas-to-dust 					&  500 \citep{Gordon:2014tm}  \\
log $\tau_{0.55}$ 			 	& 0.56$\pm$0.1 \\
 \hline
 {\bf {\sc SMMOL}}  \\
$R_{*} $ (cm)					& $1.21\times10^{14}$    (calculated from the luminosity and $T_*$)   \\
$R_{\rm in, molecules}$ (cm)	& $1.21\times10^{14}$  ($=R_{*}$)    \\
$R_{\rm in, dust}$ (cm)			&  $2.07\times10^{16}$ (adopted from the {\sc Dusty} output)      \\
$R_{\rm out}$ (cm)				& $2.07\times10^{18}$     \\
$\beta$ : $\rho\propto(r/R_{inner})^{-\beta}$ 						& 2         \\
$\alpha$: $T\propto	(r/R_{inner})^{-\alpha}$					& 0.25    (@$R<R_{\rm in, dust}$) \\
            					& 0.6$\pm$0.2    (@$R>R_{\rm in, dust}$) \\
$v_{\rm turb}$ (km\,s$^{-1}$)	&   1.0   \\
number of CO/H$_2$				& $2.7\times10^{-4}$    \\
log $\tau_{0.55}$					& 1.56 (adopted from the {\sc Dusty} output) \\
\gmas (\mlu)					&  $3\times10^{-4}$ (consistent to the {\sc Dusty} output) \\
gas-to-dust						&  500   \citep{Gordon:2014tm} \\
$v_{\rm term}$ (km\,s$^{-1}$)	& 17  \citep{Marshall:2004p9396}  \\
$v_{\rm in}$ (km\,s$^{-1}$)   	& 4.0  \\
$R_{\rm in, velocity}$ (cm)		& $1.283\times10^{15}$    \\
$\gamma$  (velocity gradient)						&  0.2     \\
 $T_{\rm sub}$ (K)  & 20    \\
\hline 
\end{tabular}\\
$T$: temperature,
 $R$: radius, 
 $\tau_{0.55}$ : optical depth at 0.55\,$\mu$m,
 \gmas: gas mass-loss rate,
 $\beta$: density law index, $(r/R_{inner})^{-\beta}$,  \newline
 $\alpha$: kinetic temperature law index, $(r/R_{inner})^{-\alpha}$, \newline
 $v_{\rm term}$: wind terminal velocity,
 $v_{\rm in}$: wind inner velocity,
 $R_{\rm in, velocity}$: radius where the wind velocity of $v_{\rm in}$  starts,
 $\gamma$: velocity law index, $v(r) = v_{inner} + (v_{\infty} - v_{inner})$  $(1- R_{inner, dust}/r)^{\gamma} $, for $r \geq R_{inner, dust}$, \newline
 and 
  $T_{\rm sub}$: sublimation temperature of CO.  
\end{center}
\end{table}

\subsection{Modelling results}

Our CO modelling results for \iras\, are displayed in Fig.\,\ref{fig-fts-sled}.
The measured CO line intensities are well fitted, particularly $J$=11--10 to 15--14.
By modelling CO emission lines, CO mass-loss rate is determined, which is used
to estimate gas mass-loss rate, assuming CO/H$_2$ abundance ratio.
We found that the estimated mass-loss rates of \iras\, from the dust modelling to the SED
and from modelling to CO emission
are consistent.

The  metallicity affects both the CO/H$_2$ abundance ratio and dust-to-gas ratio in a similar manner.
Both ratios are essentially scale with the metallicity, represented by [Fe/H].
The gas-to-dust ratio accounts for the available refractory elemental abundance.
Assuming that silicates are the major dust component in oxygen-rich circumstellar environment, such as the RSGs,
and that the silicon abundance limits the total dust mass, the estimated gas-to-dust mass ratio would be about 450.
We adopted \citeauthor{Gordon:2014tm}'s (2014) estimate of 500, which is consistent with the above estimate. 
Similarly,  the CO-H$_2$ abundance ratio accounts for the number of carbon atoms, and this parameter also scale to [Fe/H].
Essentially, the values of `dust-to-CO ratio' are the same both with solar  and LMC metallicities.


\begin{table*}
  \caption{ Luminosities and mass-loss rate of AGB stars and red-supergiants investigated via {\em Herschel} CO SLEDs \label{table-luminosities}}
\begin{center}
 \begin{tabular}{lllll}
\hline 
Name		&	Luminosity  			&  Mass-loss rate					& v$_{\rm{exp}}$ & Ref	\\ 
			&	($L_\odot$) 			& ($M_\odot$\,yr$^{-1}$)			& (km\,s$^{-1}$)			\\ \hline
\iras		&	$2.2\times10^5$		& $\sim10^{-3}$ (IR)				&	17				&	\citet{Boyer:2010ht}, \citet{Marshall:2004p9396} \\
			&							&	$3\times10^{-4}$ (IR+CO)		&					&	This work \\
WOH\,G64	&	$2.8\pm0.4\times10^5$	& $2.3\times10^{-5}$ (IR)			&	24				& \citet{Ohnaka:2008p19939},  \citet{vanLoon:1999wl}, \cite{Marshall:2004p9396} \\
VY\,CMa		&	$\sim5\times10^5$		&	$2\times10^{-4}$ (CO)			&	32 				&	\citet{Smith:2001ea}, \citet{Choi:2008p27073}, \citet{DeBeck:2010fga}, \cite{Richards:1998p27089} \\
W Hya		& 	5400					& $1.5\times10^{-7}$ (CO)			&	3--6 			& \citet{Khouri:2014gz}, \citet{Khouri:2014de},  \cite{Loup:1993p9062}\\
\hline
\end{tabular}\\
\end{center}
\end{table*}

\section{Discussions}

\subsection{CO emission}

With the {\em Herschel} SPIRE and PACS spectrometers, 
we have detected CO emission from \iras\, and WOH\,G64,
with multiple CO detections from \iras.
We modelled the \iras\, SED with a dust radiative transfer code, constraining the mass-loss rate
and the structure of the circumstellar envelope. These dust model parameters
were used to further model the CO rotational lines, resulting in good fits to the lines.

 Although the LMC has approximately
half the solar metallicity \citep{Cole:2005p839}, there is no obvious metallicity effect on
CO line intensities between Galactic and LMC red-supergiants.
Figure\,\ref{fig-fts-sled} shows the measured and modelled CO SLEDs
of the Galactic red-supergiant VY\,CMa  \citep{Matsuura:2013cz}
and the Galactic AGB star  W\,Hya \citep{Khouri:2014gz}.
The overall shapes of the CO SLEDs are alike for all three stars,
showing a sharp increase of CO at $J$=1--0 to 4--3, and flattening out at higher $J$.
No clear evidence of reduced CO intensity at lower metallicity is found in our limited sample.

\citet{Lagadec:2010kg} reported the detection of CO rotational lines from carbon-rich AGB stars, 
which are likely to be linked with the Sagittarius Dwarf Spheroidal Galaxy (Sgr\,dSph) stream.
The Sgr\,dSph has a metallicity of  [Fe/H]$\sim-1.1$   \citep{2000glg..book.....V}.
They concluded that there is no reduction in the gas mass-loss rates from carbon-rich AGB stars at this low metallicity.
In the case of carbon-rich AGB stars, the main composition of the dust is carbon, and carbon has been
 synthesised  and dredged up to the surface 
\citep[e.g.][]{Karakas:2009p25501}, and radiation pressure on carbonaceous dust grains can trigger mass loss.
Hence, it is not totally surprising that no correlation was found between mass-loss rates and metallicities for
carbon-rich AGB stars.

By contrast, the dust mass in oxygen-rich AGB stars and RSGs is expected to be restricted by the number of silicon atoms.
The silicon abundance is intrinsic to the star, i.e., limited by the silicon atoms incorporated the stars at the time of stellar formation,
although silicon might be synthesised during the very last few hundred years of the RSG phase \citep{Weaver:1980go}.
Reduced mass-loss rates are predicted at low metallicities for oxygen-rich AGB stars and RSGs
 \citep{Bowen:1991p25238}.
Therefore, it is more surprising to see no metallicity effects on the mass-loss rates for  RSGs than it is for carbon-rich AGB stars.

Instead,  we found that the CO  line intensities and the mass-loss rate
largely depends on the luminosity/luminosity class. The AGB star, W\,Hya,
which has about a factor of 50--100 lower luminosity than VY\,CMa and \iras\,
(Table\,\ref{table-luminosities}),
has about a factor of 1000 weaker CO ladder than the other two.
The luminosity appears to play the dominant role in CO line intensities and mass-loss rates.

\subsection{Mass-loss rates and luminosities at low metallicity}
 
We found from our small sample that CO line intensities and  gas mass-loss rates tend to increase with higher luminosities.
In order to further investigate the impact of luminosity, as well as metallicity, on the mass-loss rate
of evolved stars, we compare here the mass-loss rates of these stars with that of a larger sample.

The sample of Galactic stars was taken from \citet{DeBeck:2010fga}
with 47 AGB stars and RSGs.
We uses this sample because they 
used both  infrared SED and CO  lines 
in order to derive the mass-loss rates.
The Luminosities of Galactic stars were estimated from a period-luminosity relation.
For LMC stars, the sample of the 172  oxygen-rich stars from \citet{Groenewegen:2009p27401} was used,
where the mass-loss rate estimates were obtained by infrared SED modelling.
They assumed an expansion velocity of 10\,km\,s$^{-1}$ and a gas-to-dust ratio of 200 for all AGB stars and RSGs.
These assumptions can provide a factor of a few uncertainty in their mass-loss rates,
but do not have a critical impact on our systematic analysis.

\begin{figure}
\centering
   \resizebox{\hsize}{!}{\includegraphics{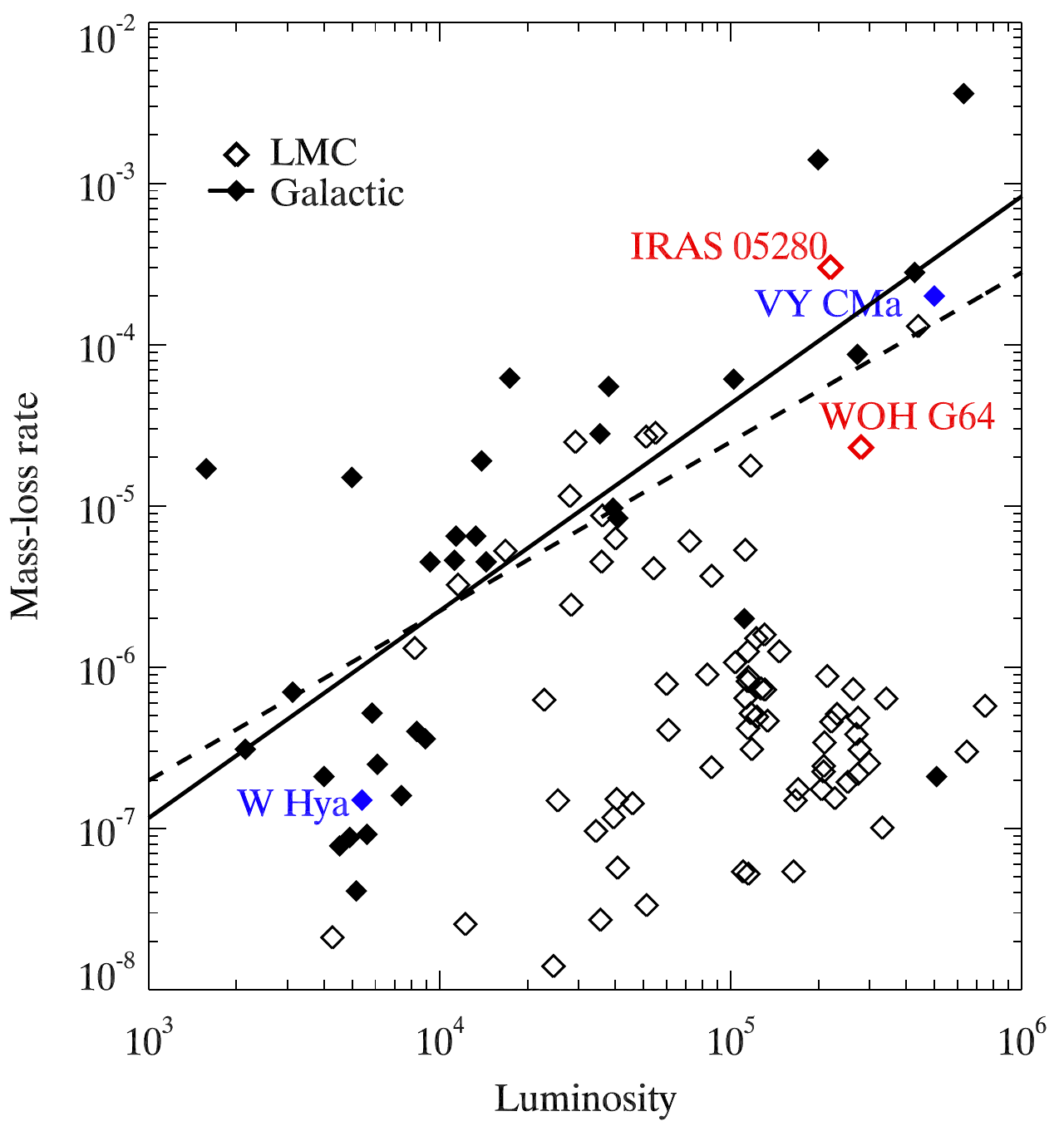}}
\caption{The mass-loss rate (\mlu) vs luminosity ($L_\odot$) 
for oxygen-rich AGB stars and red supergiants in the Galaxy and the LMC.
The Galactic data are taken from \citet{DeBeck:2010fga},
and the LMC data are from \citet{Groenewegen:2009p27401}.
The solid line shows the mass-loss vs luminosity relation fitted to Galactic sample,
and the dashed line shows the mass-loss vs luminosity relation for $T_{eff}=$3500\,K stars in the LMC
 \citep{vanLoon:2005p278}.
\label{fig-llum-mass}}
\end{figure}

 Figure\,\ref{fig-llum-mass} shows the mass-loss rate vs. luminosity relation for the LMC and Galactic evolved stars.
 The Galactic evolved stars follow an increasing trend of mass-loss rate with higher luminosity.
 This trend was fitted  with a `robust' least deviation method 
 by removing outliners \citep{1992nrfa.book.....P},
yielding
\begin{equation}
  log (\gmas) = -10.79 + 1.29\, log (L_*).
\end{equation}
The fitted line is indicated in Fig.\,\ref{fig-llum-mass}.
Both VY\,CMa and W\,Hya, which have {\em Herschel} CO SLEDs, follow this line.

In contrast, the LMC sample scatters across Fig.\,\ref{fig-llum-mass}.
Although the majority of LMC stars are placed lower than the line fitted to the Galactic objects,
two stars we studied (\iras\, and WOH\,G64) follow this Galactic trend.
Further, these two stars have among the highest mass-loss rates in the LMC.
Actually, the Galactic line approximately captures the upper range of mass-loss rates 
at a given luminosity amongst the LMC sample.

The fitted mass-loss rate vs. luminosity relation for Galactic stars is similar to that found by
\citet{vanLoon:2005p278} for LMC stars. 
Their analysis of 23 stars in the LMC produced the relation plotted as a dashed line in Fig.\ref{fig-llum-mass}. 
The line plotted in the figure is for $T_{eff}$=3500\,K. For a $T_{eff}$=3000\,K star, 
the mass-loss rate is  a factor of 2.6 higher at a given luminosity, which is small compared with scatter in the sample.
\citet{vanLoon:2005p278}'s LMC fit follows a similar trend to that found here for the Galactic stars, 
but this fit indicates only the upper range of mass-loss rate at a given luminosity 
for the larger LMC sample \citep{Groenewegen:2009p27401}.

The question is why LMC evolved stars show a much larger scatter of mass-loss rate at a given luminosity than Galactic objects.
We argue that this is due to biases in the samples. Both the LMC and Galactic stars could have a large scatter,  
but Galactic sample tend to have picked up bright CO objects.
\citet{DeBeck:2010fga}'s study was optimised for  verifying models for circumstellar envelopes with firm CO detections, 
rather than optimising for a volume-limited CO line survey.
The majority of the CO sample of \citet{DeBeck:2010fga} were taken at JCMT or APEX \citep{Kemper:2003kt}.
As \citet{Kemper:2003kt}  selected bright samples from \citet{Loup:1993p9062}'s CO catalogue, it is possible that 
mass-loss stars with weak CO lines have not been included in their target list. 
Hence, the Galactic sample could appear at the an upper end of the mass-loss vs. luminosity relation.

 As there are no systematic LMC studies of CO  emission beyond our 2 objects, 
 studies of mass-loss rates in LMC evolved stars have relied on infrared SEDs.
The LMC sample here is from \citet{Groenewegen:2009p27401}, who assembled {\em Spitzer} 
IRS spectra from six different observing programmes. 
The targets of these programmes were selected mostly from the 2MASS near-infrared survey \citep{Skrutskie:2006p2228}, 
or from a combination of 2MASS and MSX \citep{Egan:2001p216}, 
with the  exception of a few objects with optical light curves \citep{Sloan:2008p23439}.
 The near-infrared and optical surveys are efficient for detecting evolved stars 
with  relatively low mass-loss rates, but  inefficient of detecting dust-embedded AGB stars and RSGs.
For example, \iras\, does not have a 2MASS counterpart.
The surveys detected a range of stars with mass-loss rate below  $>10^{-5}$\,\mlu\, 
but only a few stars with $L>10^5$\,$L_{\odot}$, and mass-loss rate $>10^{-6}$\,\mlu\, 
was found in \citeauthor{Groenewegen:2009p27401}'s (2009) sample.
 We argue that due to the selection methods, \citeauthor{Groenewegen:2009p27401}'s (2009) LMC 
 sample  does not contain  RSGs with high optical depth dust shells.

The reason for the wide scatter in mass-loss rate and luminosity could be very interesting.
One of the possibilities is that the AGB stars that experience the third-dredge up
have an enhanced mass-loss rate  \citep[e.g.][]{Vassiliadis:1993p3075}, reaching the upper range of the mass-loss rate
and luminosity relation.

 \citet{Mauron:2011jg}  reported
a metallicity dependence for IR-SED mass loss rates,
using RSGs in the Large and Small Magellanic Clouds \citep{Bonanos:2010jk}.
The mass-loss rates would be scaled by the metallicity  $(Z/Z_{\odot})^{0.7}$.
Their definition of mass-loss rate was based on $K_s-[12]$, a good indicator of dust mass-loss rate.
Such a metallicity dependence was not detected from a comparison on three {\em Herschel} CO SLEDs
nor from a comparison between the results by  \citet{Groenewegen:2009p27401} and \citet{DeBeck:2010fga}.
Detecting metallicity effects could require a much larger CO sample.

\subsection{The contribution of high mass-loss red-supergiants to the global dust and gas source in the LMC}

 Evolved stars, both AGB stars and RSGs are considered to be important sources of dust and gas of the ISM
  \citep{Gehrz:1989p4955, Dwek:1998js, 2010pcim.book.....T}
After {\em Spitzer}'s launch, infrared colours and magnitudes were used to estimate 
individual mass-loss rates of evolved stars, and global gas and dust inputs
from evolved stars into the ISM  estimated for studied
in the Large and Small Magellanic Clouds
\citep[e.g. ][]{Matsuura:2009p29906, Srinivasan:2009p22968}.
These  studies found that a few high mass-loss rate stars are dominate the global inputs
of dust and gas from evolved stars. 
The mass loss rate of \iras\, is at the high-end range even amongst RSGs, and it must be an important source of gas and dust.

\citet{Matsuura:2013js} used 2MASS $K_s$ magnitudes for object classifications for high mass-loss rate RSGs,
but \iras\, does not have 2MASS measurements.
Similarly, \citet{Boyer:2012ck} and \citet{Riebel:2012eq} also used 2MASS photometry for their target selection.
So \iras\, was excluded from the accounting of the global gas and dust inputs into the ISM
in these two studies.

High mass-loss rate stars, such as \iras\, can contribute a significant
fraction of the total dust and gas input from evolved stars into the ISM.
\citet{Matsuura:2013js} estimated the total gas input rate from AGB stars and RSGs
to be $1.5\times10^{-2}$\,\mlu.
With a gas mass-loss rate of $2\times10^{-4}$\,\mlu,
\iras\, contribute 2\,\% of the total gas input from LMC evolved stars.
\citet{Boyer:2012ck} estimated the total `dust' input rate from AGB stars and RSGs to be
to be $1.4\times10^{-5}$\,\mlu.
The dust mass-loss rate of \iras\, is estimated to be $5\times10^{-7}$\,\mlu,
contributing about 4\,\% of the total dust input from evolved stars.
 If reasonable fractions of dust grains from RSGs may survive subsequent supernova explosions,
high mass-loss rate RSGs can  account for a substantial fraction of input to the global gas and dust budget of the LMC.

 The dust contributions from  RSGs to the ISM dust are subject to subsequent
 dust destruction by supernova (SN) shock waves, and that is largely uncertain.
Theoretical estimated values of the dust survival rate depends on models,
from 20--62\,\% \citep{Silvia:2010p29877}
to nearly 100 \% \citep{Nozawa:2010p29383, Bocchio:2014es}.
Turning to examples of nearby supernovae remnants with large dust mass detected
\citep[e.g.][]{Gomez:2012fm, Barlow:2010p29287, Matsuura:2015kn}, 
both Cassiopeia A and Supernova 1987A show the presence of reverse shocks
 \citep{Gotthelf:2001bn, France:2010p29071},
and dust was detected from the unshocked regions 
\citep{Barlow:2010p29287, Indebetouw:2014bt}.
In contrast, the dominant energy in the Crab Nebula is currently from its pulsar wind nebula.
In the older supernova remnant, the Cygnus Loop, an enhanced C abundance in shocked region suggests 
the destruction of carbonaceous dust
\citet{Raymond:2013en}, but this could be swept-up ISM dust.
\citet{temim:2015bs} and \citet{Lakicevic:2015iw} studied supernova remnants (SNRs) in the Large Magellanic Clouds,
and estimated that dust overall destroyed by supernovae. \citet{Lakicevic:2015iw}  found that
the temperature of SNR dust emission to be higher than that of ISM dust, which affect their dust mass analysis.
In the optically thin case, the thermal emission from warmer SN dust can dominate the total emission,  
hiding weaker emission from colder ISM dust.
This effect  distort the measured ISM dust mass towards the SNRs
rather than the ISM dust mass being actually lower towards the SNRs due to
SN shocks destroying surrounding ISM dust grains.
A challenge remains in measuring dust mass destroyed by supernovae.


\section{Summary}


We have detected far-infrared and sub-millimetre  CO  and H$_2$O lines from two red-supergiants in the LMC.
These are the first detections of these lines in LMC evolved stars.

Modelling with radiative transfer codes of the dust and line emission resulted in good fits to the infrared-SED and 
CO transitions of \iras.
A mass-loss rate of $3\times10^{-4}$\,\mlu\, was obtained,
and a value is at the high end of mass-loss rates found for red-supergiants.
A gas-to-dust ratio of 500 and a CO/H$_2$ abundance of $2.7\times10^{-4}$ were adopted for the modelling.
The gas-to-dust ratio was adopted from the value estimated for the LMC ISM, but is consistent with 
the solar neighbourhood value scaled by the LMC metallicity.
The CO/H$_2$ abundance  was adopted from the Galactic value scaled by the LMC metallicity.
Our good modelling results support  CO/dust ratio appears to be constant in the Galactic and LMC RSGs.

There is a general increasing trend of mass-loss rate with higher luminosity.
The LMC red-supergiants with CO detections appear at the  an upper range of  mass-loss rate at a given luminosity. 
Among CO-detected AGB stars and red-supergiants, there is no obvious metallicity effect found between Galactic
and LMC  stars and object-to-object variation overwhelms the metallicity dependence, if any.

\section{acknowledgments}
 We would like to thank S. Srinivasan for useful inputs on $\chi^2$ analysis on {\sc Dusty} fitting.
M.M. is supported by the STFC Ernest Rutherford fellowship.
M. Meixner  and B.S. acknowledge funding from the NASA Astrophysics Data Analysis Program 
grant NNX13AD54G and from the NASA/JPL grant NNN12AA01C.
PACS has been developed by a consortium of institutes led by MPE (Germany) and including UVIE (Austria); KU Leuven, CSL, IMEC (Belgium); CEA, LAM (France); MPIA (Germany); INAF-IFSI/OAA/OAP/OAT, LENS, SISSA (Italy); IAC (Spain). This development has been supported by the funding agencies BMVIT (Austria), ESA-PRODEX (Belgium), CEA/CNES (France), DLR (Germany), ASI/INAF (Italy), and CICYT/MCYT (Spain).
SPIRE has been developed by a consortium of institutes led by Cardiff University (UK) and including Univ. Lethbridge (Canada); NAOC (China); CEA, LAM (France); IFSI, Univ. Padua (Italy); IAC (Spain); Stockholm Observatory (Sweden); Imperial College London, RAL, UCL-MSSL, UKATC, Univ. Sussex (UK); and Caltech, JPL, NHSC, Univ. Colorado (USA). This development has been supported by national funding agencies: CSA (Canada); NAOC (China); CEA, CNES, CNRS (France); ASI (Italy); MCINN (Spain); SNSB (Sweden); STFC and UKSA (UK); and NASA (USA).
The VISTA magnitude is based on observations collected at the European Organisation for Astronomical Research in the Southern Hemisphere under ESO programme(s) 179.B-2003.

\bibliography{v10}

\bibliographystyle{mn2e}


\label{lastpage}

\end{document}